\documentclass[amsmath, amssymb, aps, prb, nofootinbib, longbibliography, reprint]{revtex4-2}
\usepackage{booktabs}
\usepackage{footnote}
\usepackage{graphicx}
\usepackage{dcolumn}
\usepackage{bm}

\renewcommand{\selectlanguage}[1]{}

\begin{document}

\title{\textit{Ab~initio} modeling of nonequilibrium dynamics in superconducting detectors and qubits}

\author{Alejandro Simon}
\email{Corresponding author: alejansi@mit.edu}
\affiliation{%
 Department of Electrical Engineering and Computer Science, Massachusetts Institute of Technology \\
 50 Vassar Street, Cambridge, MA, USA, 02139-4307
}%
\author{Reed Foster}%
\affiliation{%
 Department of Electrical Engineering and Computer Science, Massachusetts Institute of Technology \\
 50 Vassar Street, Cambridge, MA, USA, 02139-4307
}%

\author{Mihir Sahoo}
\affiliation{Institute of Theoretical and Computational Physics, Graz University of Technology \\ 
Petersgasse 16, 8010 Graz, Austria}

\author{James Shi}
\affiliation{%
 Department of Electrical Engineering and Computer Science, Massachusetts Institute of Technology \\
 50 Vassar Street, Cambridge, MA, USA, 02139-4307
}%

\author{Emma Batson}
\affiliation{%
 Department of Electrical Engineering and Computer Science, Massachusetts Institute of Technology \\
 50 Vassar Street, Cambridge, MA, USA, 02139-4307
}%
\author{Francesca Incalza}
\affiliation{%
 Department of Electrical Engineering and Computer Science, Massachusetts Institute of Technology \\
 50 Vassar Street, Cambridge, MA, USA, 02139-4307
}%

\author{Matteo Castellani}
\affiliation{%
 Department of Electrical Engineering and Computer Science, Massachusetts Institute of Technology \\
 50 Vassar Street, Cambridge, MA, USA, 02139-4307
}%

\author{Owen Medeiros}
\affiliation{%
 Department of Electrical Engineering and Computer Science, Massachusetts Institute of Technology \\
 50 Vassar Street, Cambridge, MA, USA, 02139-4307
}%

\author{Christoph Heil}
\affiliation{Institute of Theoretical and Computational Physics, Graz University of Technology \\ 
Petersgasse 16, 8010 Graz, Austria}

\author{Karl K. Berggren}
\affiliation{%
 Department of Electrical Engineering and Computer Science, Massachusetts Institute of Technology \\
 50 Vassar Street, Cambridge, MA, USA, 02139-4307
}%

\date{\today}

\begin{abstract}
Nonequilibrium quasiparticle and phonon dynamics are central to the operation of superconducting devices. Superconducting detectors, such as superconducting nanowire single-photon detectors, transition-edge sensors, or microwave kinetic inductance detectors, perform best when a large quasiparticle population is generated in response to small perturbations. Conversely, for superconducting qubits and topologically protected Majorana fermions, even relatively small quasiparticle densities can lead to significant performance degradation. Hence, ideal materials for these devices would be less susceptible to quasiparticle poisoning. However, existing models of these devices often rely on approximations and phenomenology. Therefore, they lack a rigorous description of the underlying quasiparticle and phonon dynamics that are responsible for device performance. In this article, we combine kinetic equations with density functional theory to model the non-equilibrium quasiparticle and phonon dynamics of a thin superconducting film \textit{ab initio}. To demonstrate the universality of our model, we illustrate two independent example applications: (1) we develop a theoretical model describing the detection of single photons in superconducting nanowires, and (2) we calculate the energy-relaxation rate of a transmon qubit due to the presence of excess quasiparticles. Our examples demonstrate from first principles that niobium nitride is well-suited to be used for single-photon detection and that tantalum transmon qubits possess reduced sensitivity to quasiparticle poisoning relative to other materials, which is likely in part responsible for their longer coherence times. In contrast to previous models of superconducting devices, our \textit{ab initio} approach makes predictions of device performance without experimental input and thus can be used to accelerate progress in device development. Moreover, by considering the full-bandwidth electron-phonon coupling, our approach can incorporate strong-coupling effects. Our methods effectively integrate \textit{ab initio} materials modeling with nonequilibrium theory of superconductivity to perform practical modeling of superconducting devices, providing a comprehensive approach that connects fundamental theory with device-level applications. 
\end{abstract}

\maketitle

\section{Introduction}

The success of the Migdal-Eliashberg theory has illustrated the importance of considering the full-bandwidth electron-phonon coupling spectrum to describe conventional superconductivity \cite{eliashberg1960interactions, PhysRevB.12.905, Pellegrini2024}. However, the full-bandwidth electron-phonon coupling not only provides insights into equilibrium superconductivity but also lays the foundation for understanding nonequilibrium phenomena, where the electron-phonon coupling determines the evolution of the quasiparticle and phonon distributions. These distributions, in turn, influence the modification of the transport properties of a material when subjected to external perturbations, such as radiation absorption or heating \cite{prange_transport_1964, eliashberg1970film, 
eliashberg1972inelastic, 
chang_kinetic-equation_1977}. However, obtaining the full-bandwidth electron-phonon coupling spectrum for arbitrary materials has historically been difficult. Thus, studies of the nonequilibrium dynamics of superconductors typically rely on approximating the phonon system with a Debye model \cite{
debye1912theorie, 
chang_kinetic-equation_1977, ashby_detailed_1981, goldie2012non}. This approximation, which assumes a linear dispersion for the phonons, is generally inadequate for capturing realistic electron-phonon coupling. Therefore, the Debye model can only provide qualitative predictions and cannot be used to describe general nonequilibrium superconductivity. 

This limitation of the Debye model is a critical issue for device modeling, as nonequilibrium dynamics are central to the operation of many superconducting devices. To overcome this limitation, we developed an \textit{ab initio} model to quantitatively predict the microscopic response of a thin, narrow superconducting film to external perturbation. To do so, we employ recent advances in \textit{ab initio} materials modeling within the framework of Density Functional Perturbation Theory (DFPT) \cite{Lucrezi2024, lee2023electron}. The resulting model is of particular interest for the development and study of superconducting detectors \cite{vodolazov_single-photon_2017, allmaras2020modeling}, microwave resonators \cite{budoyo2016effects}, and quasiparticle poisoning of qubits \cite{PhysRevLett.132.017001, PhysRevLett.133.060602}. 


In the following, we formulate the theory behind this model and apply it in two independent examples: (1) to describe a film irradiated by optical photons and (2) to describe the energy-relaxation rate of a transmon qubit. To connect our first example with experiment, we use our model to predict the wavelength sensitivity of a superconducting nanowire single-photon detector (SNSPD) by determining the detection current $I_{\mathrm{det}}$, defined as the current at which the internal detection efficiency of an SNSPD saturates for a given photon wavelength and device temperature, and compare our results to experimental data. For our second example, we study the energy-relaxation rate of the transmon when impacted by ionizing radiation. For comparison with experimental results, we focus on niobium nitride ($\mathrm{NbN}$) when modeling optical irradiation and aluminum (Al), niobium (Nb), titanium nitride (TiN), lead (Pb), and tantalum (Ta) when modeling transmon energy-relaxation rates; however, the methods outlined here are suitable to describe any conventional isotropic superconductor and can be generalized to incorporate anisotropy \cite{prange_transport_1964}. We also emphasize that as an \textit{ab initio} theory, the predictions of this model are based on first-principles calculations of the material's properties and can be made with no experimental input. 

\begin{figure*}[t]
    \centering
    \includegraphics[width=\linewidth]{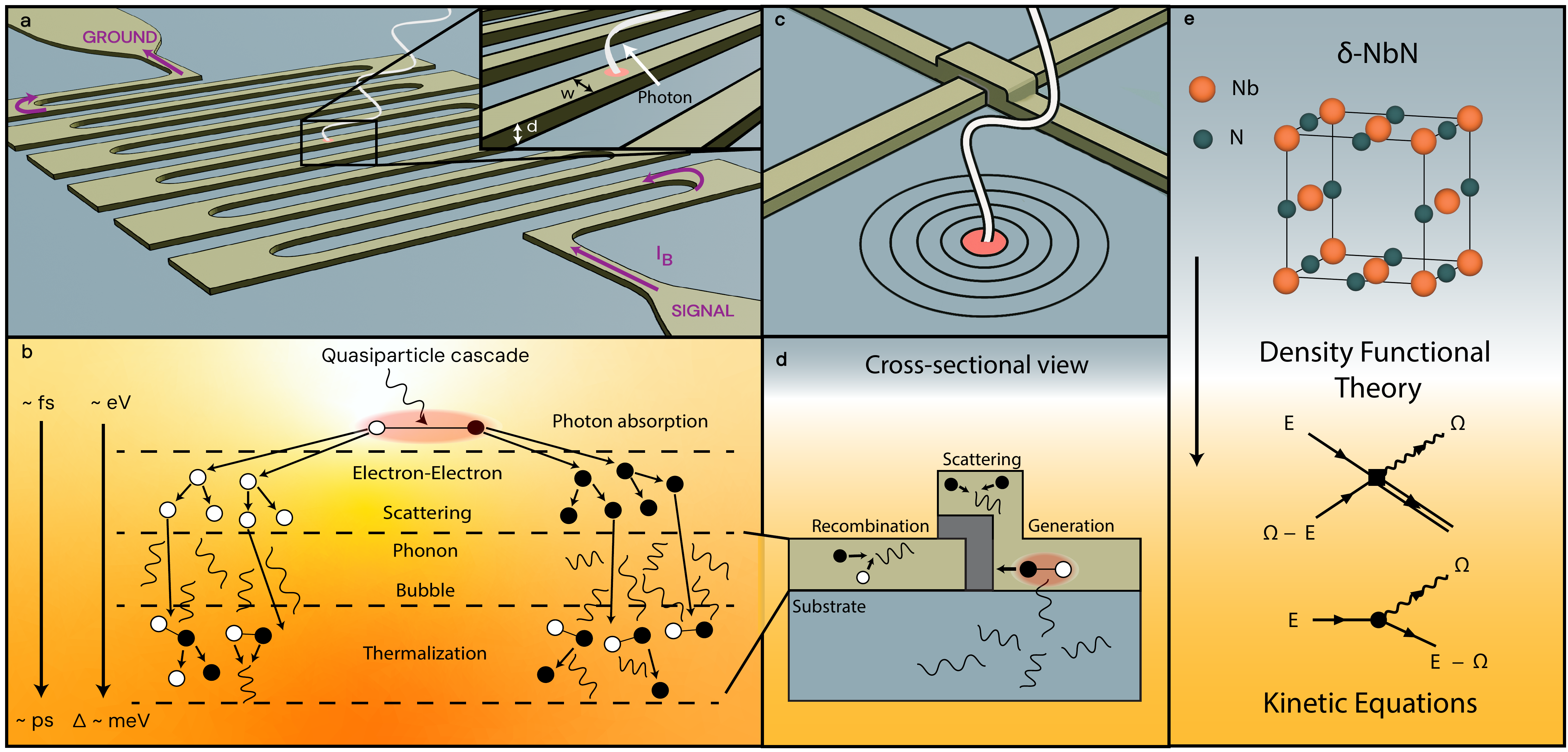}
    \caption{High-level overview of the modeling of nonequilibrium devices. (a) Typical SNSPD geometry consisting of a thin ($d \sim \xi_{\mathrm{c}}$) narrow ($w \ll \Lambda$) superconducting wire that is patterned in a meander to increase the active area of the detector. The device is single-photon sensitive when a sufficiently large bias current $I_{\mathrm{B}}$ is applied. (b) Microscopic picture of SNSPD detection. A photon is absorbed, generating an excited quasiparticle consisting of an electron-hole pair. This excitation triggers an energy-relaxation cascade and the generation of a phonon bubble. The resulting quasiparticles and phonons scatter and break pairs, locally suppressing the superconductivity and weakening the barrier for quantum and thermal fluctuations that may fully destroy the superconductivity across the strip. Due to the bias current, this normal strip produces a nonzero voltage across the terminals of the device, which is read out as a detection event. (c) A cartoon depiction of the Manhattan-style Josephson junction commonly used in transmon qubit fabrication. A cosmic ray interacts with the substrate, generating a large population of phonons that perturb the superconductivity in the Josephson junction. (d) Cross-sectional view of the process in (c). The quasiparticle recombination, scattering, and generation processes are displayed inside the superconducting layer of the Josephson junction. These processes are identical to those in the phonon bubble and thermalization steps shown in (b). (e) \textit{Ab initio} approach to modeling nonequilibrium dynamics in superconductors. Beginning with the crystal structure, we obtain the electron-phonon coupling with density functional theory, which allows us to obtain the interaction probabilities for the quasiparticle and phonon systems. The evolution of the quasiparticle and phonon distributions is then modeled using a set of kinetic equations.}
    \label{fig:high-level}
\end{figure*}

\section{Model formulation}

\begin{figure*}
    \centering
    \includegraphics[width=\linewidth]{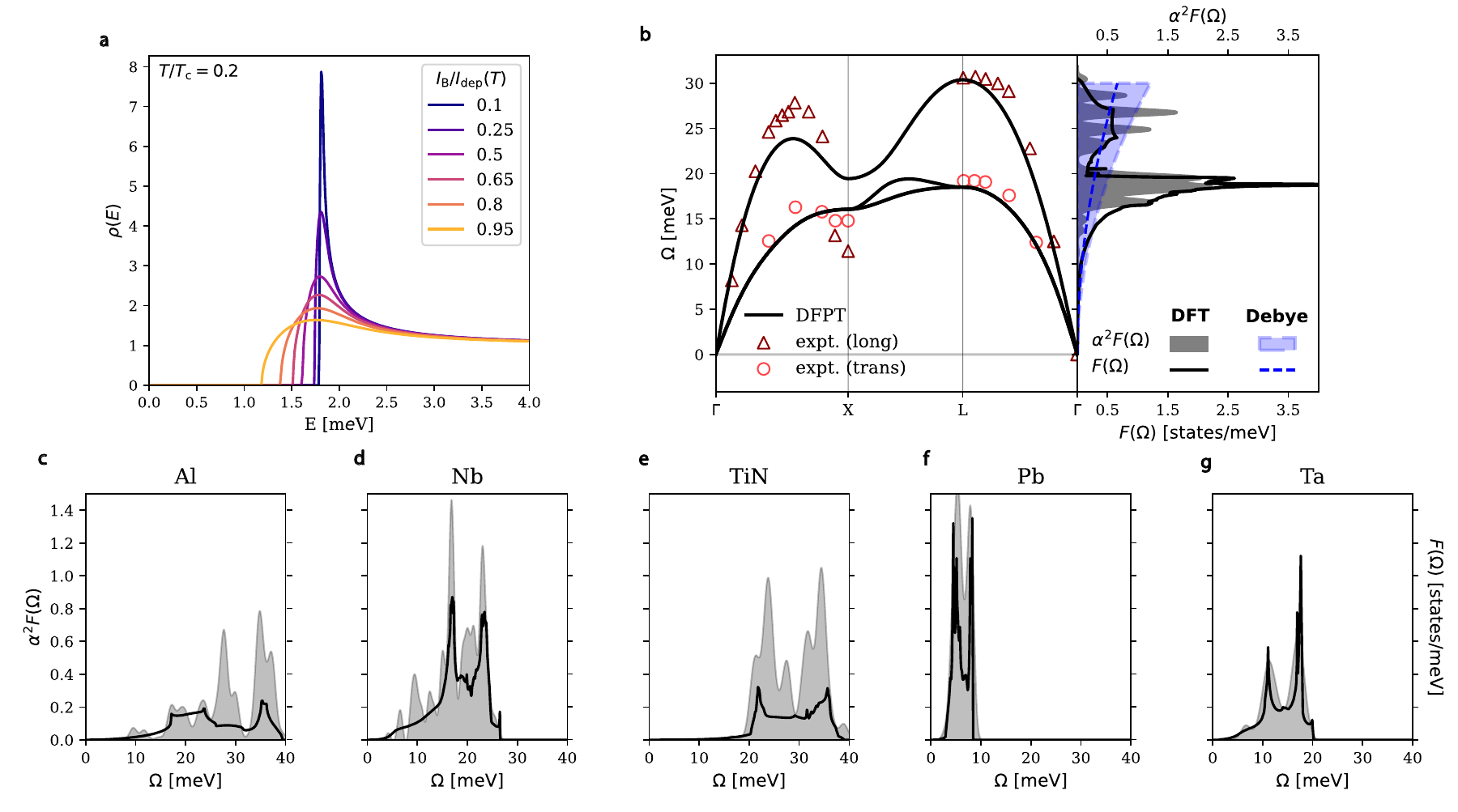}
    \caption{(a) Normalized quasiparticle density of states $\rho(E)$ for $\delta$-$\mathrm{NbN}$ with $T_{\mathrm{c}} = 10\,\mathrm{K}$ and different ratios of the bias current normalized to the depairing current $I_{\mathrm{B}}/I_{\mathrm{dep}}(T)$. (b) Acoustic phonon modes for $\delta$-$\mathrm{NbN}$ calculated using density functional perturbation theory compared to experimental data for the longitudinal and transverse modes \cite{christensen1979phonon}. The corresponding Eliashberg spectral function $\alpha^2F(\Omega)$ (solid grey) and phonon density of states $F(\Omega)$ (black) are displayed on the right compared to the Debye approximation (dashed and solid blue). Density functional perturbation theory calculated results are also displayed for (c) aluminum, (d) niobium, (e) titanium nitride, (f) lead, (g) tantalum. Spin-orbit coupling effects were included for lead and tantalum. }
    \label{fig:DFT}
\end{figure*}

When a superconductor is perturbed, excess quasiparticles and/or phonons are generated. The resulting nonequilibrium dynamics of the relaxation process can be described by a set of kinetic equations for the quasiparticle $f(E)$ and phonon $n(\Omega)$ distribution functions, which for an isotropic material and neglecting diffusion are
\begin{subequations}
\begin{equation}
\label{eq:qp-dist}
\begin{aligned}
    \frac{d f(E)}{d t} = & -\frac{2 \pi}{\hbar} \int_{0}^{\infty} d\Omega \, \alpha^2 F(\Omega) \rho(E+\Omega) K_{\mathrm{ph}-e}(E, \Omega)
    \\ & -\frac{2 \pi}{\hbar} \int_0^{E-|\Delta|} d\Omega \, \alpha^2F(\Omega) \rho(E-\Omega) K_{e-\mathrm{ph}}(E, \Omega)
    \\ & -\frac{2 \pi}{\hbar} \int_{E+|\Delta|}^{\infty} d\Omega \, \alpha^2 F(\Omega) \rho(\Omega - E) K_{\mathrm{R}}(E,\Omega) 
\end{aligned}
\end{equation}
and 
\begin{equation}
\label{eq:ph-dist}
\begin{aligned}
    \frac{d n(\Omega)}{d t} &  = -\frac{8 \pi}{\hbar}  \frac{N(0)}{N} \int_{|\Delta|}^{\infty} d\!E  \int_{|\Delta|}^{\infty} d\!E' \, \alpha^2(\Omega) \rho(E) \rho(E') \\ & \times  \Bigg[ K_{\mathrm{S}}(E, E', \Omega) \delta(E+\Omega-E') \\ & + K_{\mathrm{B}}(E,E',\Omega)\delta(E+E'-\Omega) \Bigg]-\frac{n(\Omega) - n^{\mathrm{eq}}(\Omega)}{\tau_{\mathrm{esc}}},
\end{aligned}
\end{equation}
\end{subequations}
where the integral kernels $K_i$ are functions of $f(E)$ and $n(\Omega)$ and are defined in the Appendix. $K_{\mathrm{ph}-e}(E, \Omega)$ ($K_{e-\mathrm{ph}}(E, \Omega)$) describes quasiparticle scattering due to the absorption (emission) of a phonon, while $K_{\mathrm{R}}(E,\Omega)$ represents the quasiparticle recombination process. $K_{\mathrm{S}}(E, E', \Omega)$ captures the phonon scattering process, and $K_{\mathrm{B}}(E, E', \Omega)$ describes the phonon pair-breaking process. Here, $N$ is the number of ions per unit volume, and $\rho(E)$ is the normalized quasiparticle density of states \cite{chang_kinetic-equation_1977}. These equations are valid in cases where the quasiparticle approximation holds, i.e., when the quasiparticle lifetimes are sufficiently long, corresponding to well-defined states. This is true for the excitation energies of interest in this paper, which are below the characteristic phonon frequency $\omega_{\rm{D}}$ even in the case of optical irradiation. In Eqs. \eqref{eq:qp-dist} and \eqref{eq:ph-dist}, it is also assumed that renormalization effects are small and that the imaginary component of the order parameter $\Delta$ is negligible. For small excitation frequencies relative to $\omega_{\rm{D}}$, we may also neglect the frequency dependence of $\Delta$. For a film in the dirty limit, characterized by $l_{e} \ll \xi_{\mathrm{c}}$, $\rho(E)$ can be calculated for a finite bias current $I_{\mathrm{B}}$ by solving the Usadel equation as detailed in the Appendix. Solutions to the Usadel equation for $\rho(E)$ at various bias currents are displayed in Fig. \ref{fig:DFT}a. With the above simplifications, the superconducting order parameter $\Delta$ satisfies
\begin{equation}
\label{eq:real-axis-sc}
|\Delta| = \lambda\int_{|\Delta|}^{\infty} d\!E \, R(E,\Delta) [1-2f(E)] , 
\end{equation} where $\lambda = 2\int_0^{\infty} d\Omega \alpha^2F(\Omega)/\Omega$ is the electron-phonon coupling parameter and $R(E,\Delta)$ is a spectral function defined in the Appendix \cite{chang_kinetic-equation_1977,vodolazov_single-photon_2017}. The final term of Eq. \eqref{eq:ph-dist} models phonon exchange with the substrate, where $n^{\mathrm{eq}}(\Omega)$ is the usual Bose-Einstein distribution and $\tau_{\mathrm{esc}}$ is the characteristic time for phonon escape to the substrate. To simplify calculations, we ignore the energy and angular dependence of $\tau_{\mathrm{esc}}$.

In Eqs. \eqref{eq:qp-dist} and \eqref{eq:ph-dist}, the quasiparticle and phonon interaction probabilities are determined by the Eliashberg spectral function $\alpha^2 F(\Omega)$ and the phonon density of states $F(\Omega)$. In general, $\alpha^2F(\Omega)$ and $F(\Omega)$ can be obtained experimentally through electron-tunneling and inelastic neutron-scattering measurements, respectively. However, with DFPT, we can computationally obtain these quantities \textit{ab initio} for a wide range of conventional superconductors, including superconducting alloys and anisotropic materials \cite{Lucrezi2024, FERREIRA2024101547}, circumventing the need for experimental data. As we show in Fig. \ref{fig:high-level}e, in our model, we combine DFPT calculations with Eqs. \eqref{eq:qp-dist} and \eqref{eq:ph-dist} to model the nonequilibrium quasiparticle and phonon dynamics. In Fig. \ref{fig:DFT}b, we display the calculated acoustic phonon dispersion for the $\delta$-NbN phase within the harmonic approximation and experimental data obtained via neutron scattering for $\delta$-NbN$_{0.93}$ \cite{christensen1979phonon}. Notably, the agreement is strong for the acoustic branches, which exhibit the strongest electron-phonon coupling, underscoring the validity of our theoretical approach. Details of the DFPT calculations are contained in the Appendix. We also include DFPT-calculated $\alpha^2F(\Omega)$ and $F(\Omega)$ for Al, Nb, TiN, Pb, and Ta in Fig. \ref{fig:DFT}c-g. 

In Fig. \ref{fig:DFT}b, $\alpha^2 F(\Omega)$ and $F(\Omega)$ for $\mathrm{NbN}$ are displayed alongside the Debye model, where a quadratic frequency dependence is assumed with $\alpha^2 F(\Omega) = \lambda \Omega^2/\Omega_\mathrm{D}^2$ and $F(\Omega) = 9 \Omega^2 / \Omega_\mathrm{D}^3$ for $\Omega < \Omega_{\mathrm{D}}$ and zero otherwise, where $\Omega_\mathrm{D}$ is the Debye frequency \cite{chang_kinetic-equation_1977, goldie2012non}. This comparison clearly shows that the structure of $\alpha^2 F(\Omega)$ and $F(\Omega)$ is neglected in the Debye approximation. Given the importance of these quantities in determining the interaction probabilities, one must consider their precise forms to make quantitative predictions of the nonequilibrium quasiparticle and phonon dynamics. 

\section{Optical Irradiation}

Due to their exceptional performance characteristics, superconducting single-photon detectors, such as the SNSPD, transition-edge sensor, and microwave kinetic inductance detector, have quickly become ubiquitous in sensing, communication, and computing applications. In particular, SNSPDs have earned widespread recognition for demonstrations of near-unity internal single-photon detection efficiency \cite{marsili_detecting_2013, reddy_superconducting_2020, chang_detecting_2021}, single-photon sensitivity from X-rays to mid-infrared wavelengths \cite{verma_single-photon_2021, Taylor:23}, ultra-low dark-count rates \cite{PhysRevLett.123.151802, hochberg_new_2022}, and sub-$3\,\rm{ps}$ timing jitter \cite{korzh_demonstration_2020}. However, applications such as dark matter search, biomedical imaging, and space communication can benefit considerably from improvements in the operating temperature and wavelength sensitivity of these detectors. Consequently, there is a large effort directed at engineering and exploring new materials for enhanced SNSPD performance \cite{charaev_single-photon_2024, 
shibata_single-photon_2010, 
cherednichenko_low_2021, 
arpaia_high-temperature_2015, 
arpaia_transport_2017, 
ejrnaes_observation_2017, 
charaev_single-photon_2023, korzh_demonstration_2020, chang_detecting_2021, Taylor:23}. 


To direct this effort, a precise understanding of the mechanism underpinning photon detection in SNSPDs is required \cite{Semenov2001, 
PhysRevB.52.581, 
PhysRevB.79.100509, 
4277823, 
PhysRevB.84.174510, 
PhysRevB.75.094513, 
PhysRevB.61.11807,
vodolazov_current_2014, 
Engel_2015, 
vodolazov_single-photon_2017, 
Berggren_2018, 
allmaras2020modeling}. To this end, several phenomenological models of photon detection in superconducting nanowires have been proposed; however, these models are generally unsatisfactory for describing arbitrary SNSPD geometries and materials. Prior work has demonstrated the crucial role of quasiparticle and phonon interactions in the initial stages of photon detection \cite{PhysRevB.75.094513, 
PhysRevB.61.11807, 
vodolazov_single-photon_2017, 
allmaras2020modeling, 
vodolazov_current_2014, 
Semenov_2021}. Only a limited number of studies have attempted to model these interactions directly \cite{PhysRevB.75.094513, vodolazov_single-photon_2017, allmaras2020modeling}; however, these studies have relied on the Debye approximation, consequently neglecting the effect of realistic electron-phonon coupling. 


In this work, we address this limitation with the \textit{ab initio} model. We consider a typical SNSPD geometry, consisting of a thin, narrow superconducting wire that has absorbed a single optical photon while carrying a nonzero bias current $I_{\mathrm{B}}$. As depicted in Fig. \ref{fig:high-level}a, such a film is characterized by a thickness on the order of the superconducting coherence length $d \sim \xi_\mathrm{c} = \sqrt{\hbar D / |\Delta|}$ and width much smaller than the Pearl length $w \ll \Lambda = 2\lambda_\mathrm{L}^2 / d$, where $\lambda_\mathrm{L}$ is the London penetration depth ($\lambda_{\mathrm{L}} = \sqrt{\hbar \rho_{\rm{N}}/\mu_0 \pi |\Delta|}$), $\rho_{\rm{N}}$ is the normal state resistivity ($\rho_{\mathrm{N}} = 1/2 e^2 D N(0)$), $e$ is the electron charge, $N(0)$ is the single-spin electron density of states at the Fermi energy $E_\mathrm{F}$, $\mu_0$ is the permeability of free-space, $\hbar$ is the reduced Planck constant, $D$ is the electronic diffusion coefficient ($D = v_{\mathrm{F}} l_{e}/3$), $v_{\rm F}$ is the Fermi velocity, $l_{e}$ is the electron mean-free path, and $|\Delta|$ is the magnitude of the superconducting order parameter. In general, $|\Delta|$ is a function of temperature $T$, $I_{\mathrm{B}}$, and position $\mathbf{r}$. 

When a photon is absorbed in the superconducting film, a single quasiparticle is excited with an energy $E_\lambda \gg |\Delta|$. The lifetime of a quasiparticle of energy $\sim E_\lambda$ is extremely short relative to the timescale of variations of the superconducting order parameter $\tau_{\Delta} = \hbar / |\Delta|$ \cite{kaplan1976quasiparticle, 10.1063/1.122942}. Thus, the subsequent interactions are practically instantaneous from the perspective of $\Delta$. Initially, quasiparticle relaxation occurs primarily through electron-electron scattering and the emission of secondary electrons. These electrons quickly reach energies on the order of $\Omega_{\mathrm{D}} = \hbar \omega_{\rm{D}}$, where relaxation via acoustic-phonon emission dominates \cite{chang_kinetic-equation_1977, 10.1063/1.122942}. These emitted phonons possess short mean-free paths and contribute to pair-breaking. Hence, the initial distribution for Eqs. \eqref{eq:qp-dist} and \eqref{eq:ph-dist} is well approximated by a phonon-bubble initial condition \cite{vodolazov_single-photon_2017}, which is derived in the Appendix. These dynamics are illustrated in Fig. \ref{fig:high-level}b. 

\begin{figure*}
    \centering
    \includegraphics[width=\linewidth]{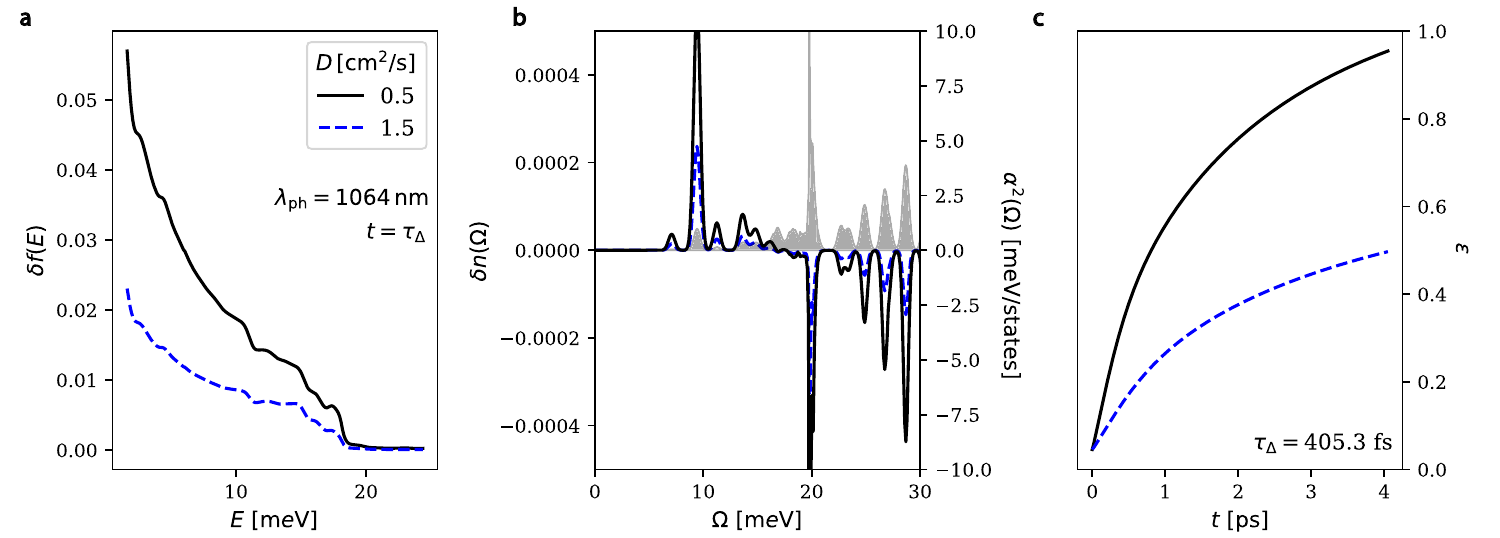}
    \caption{Solutions to the kinetic equations for $\delta f(E)$, $\delta n(\Omega)$, and $\epsilon(t)$ for NbN. (a) Nonequilibrium excess quasiparticle $\delta\!f(E)$ and (b) excess phonon $\delta \! n(\Omega)$ distribution generated by the absorption of a photon with wavelength $\lambda_{\mathrm{ph}} = 1064\,\mathrm{nm}$ at $t=\tau_{\Delta}$ for $\mathrm{NbN}$ with electronic diffusion coefficients of $D=0.5\,\mathrm{cm^2/s}$ (solid black lines) and  $D = 1.5\,\mathrm{cm^2/s}$ (dashed blue lines). In (b), $\alpha^2(\Omega)$ for NbN is displayed on the right axis. (c) Quasiparticle-induced suppression parameter $\varepsilon(t)$. A value of $\varepsilon = 0$ implies no suppression of $\Delta$.}
    \label{fig:suppression}
\end{figure*}

In Fig. \ref{fig:suppression}a and Fig. \ref{fig:suppression}b, numerical solutions to Eqs. \eqref{eq:qp-dist} and \eqref{eq:ph-dist} for two different electronic diffusion coefficients $D$ are displayed. In these calculations, material parameters consistent with $\mathrm{NbN}$, and SNSPD geometries, were used and can be found in the Appendix. $D=1.5\,\mathrm{cm^2/s}$ is typical of epitaxial $\mathrm{NbN}$ \cite{PhysRevB.77.214503}, while $D=0.5\,\mathrm{cm^2/s}$ is typical of polycrystalline $\mathrm{NbN}$ \cite{vodolazov_single-photon_2017}. For smaller $D$, corresponding to greater disorder, $\alpha^2F(\Omega)$ is smeared; however, we do not expect this effect to have a significant impact on our results, and thus, we neglect it. In our solutions, we assume that the photon's energy is initially distributed uniformly in a cylindrical volume of $V_{\mathrm{init}} = \pi \xi_{\rm{c}}^2 d$ and $|\Delta(I_{\mathrm{B}}, T)|$ is constant for the timescales of interest ($t \sim  \tau_{\Delta}$). Further details regarding the numerical methods and validation are discussed in the Appendix.

By inserting $\delta\!f(E)$ into Eq. \eqref{eq:real-axis-sc} we calculate the quasiparticle-induced suppression parameter  
\begin{equation}
    \label{eq:epsilon}
    \varepsilon(t) = 2 \int_{|\Delta|}^{\infty}d\!E\, \frac{R(E,\Delta)}{|\Delta|} \delta\! f(E),
\end{equation} which characterizes the suppression of $|\Delta|$ and is displayed in Fig. \ref{fig:suppression}c. Fig. \ref{fig:suppression}c illustrates that in dirtier materials with a smaller $D$, a larger nonequilibrium quasiparticle population is generated within $V_{\rm{init}}$, resulting in a more significant suppression of $|\Delta|$ in the initial stages following photon absorption. These results are consistent with the argument that a larger $D$ leads to more stringent requirements on the detector's geometry to maintain photon sensitivity, e.g., reducing the film thickness and/or wire width \cite{10.1063/5.0018818}.

We define the thermalization time of the quasiparticle and phonon system $\tau_{\mathrm{th}}$ by fitting the numerical solution for $\varepsilon(t)$ to an exponential $\varepsilon(t) = \varepsilon_{t \rightarrow \infty} ( 1 - e^{-t/\tau_{\mathrm{th}}} )$. For $\mathrm{NbN}$ at $T/T_{\rm{c}}=0.2$ with an electronic diffusion coefficient of $D=0.5\,\mathrm{cm^2/s}$ we find $\tau_{\mathrm{th}} = 1.4\,\mathrm{ps}$ and for $D=1.5\,\mathrm{cm^2/s}$ we find $\tau_{\mathrm{th}} = 1.8\,\mathrm{ps}$ for a photon wavelength of $\lambda_{\mathrm{ph}}=1064\,\mathrm{nm}$. That $\tau_{\mathrm{th}} > \tau_{\Delta} = 405.3\,\mathrm{fs}$ is consistent with results obtained with the Debye model \cite{vodolazov_single-photon_2017}; however, with the Debye model it is found that, for $D = 0.5\,\mathrm{cm^2/s}$, $\tau_{\mathrm{th}} \approx 1.5 \tau_{\Delta}$, whereas with the full-bandwidth electron-phonon coupling, we find $\tau_{\mathrm{th}} \approx 4 \tau_{\Delta}$. The larger value of $\tau_{\mathrm{th}}$ obtained with the full-bandwidth electron-phonon coupling is in better agreement with experimental data \cite{10.1063/1.122942}. 

\begin{figure*}
    \centering
    \includegraphics[width=\linewidth]{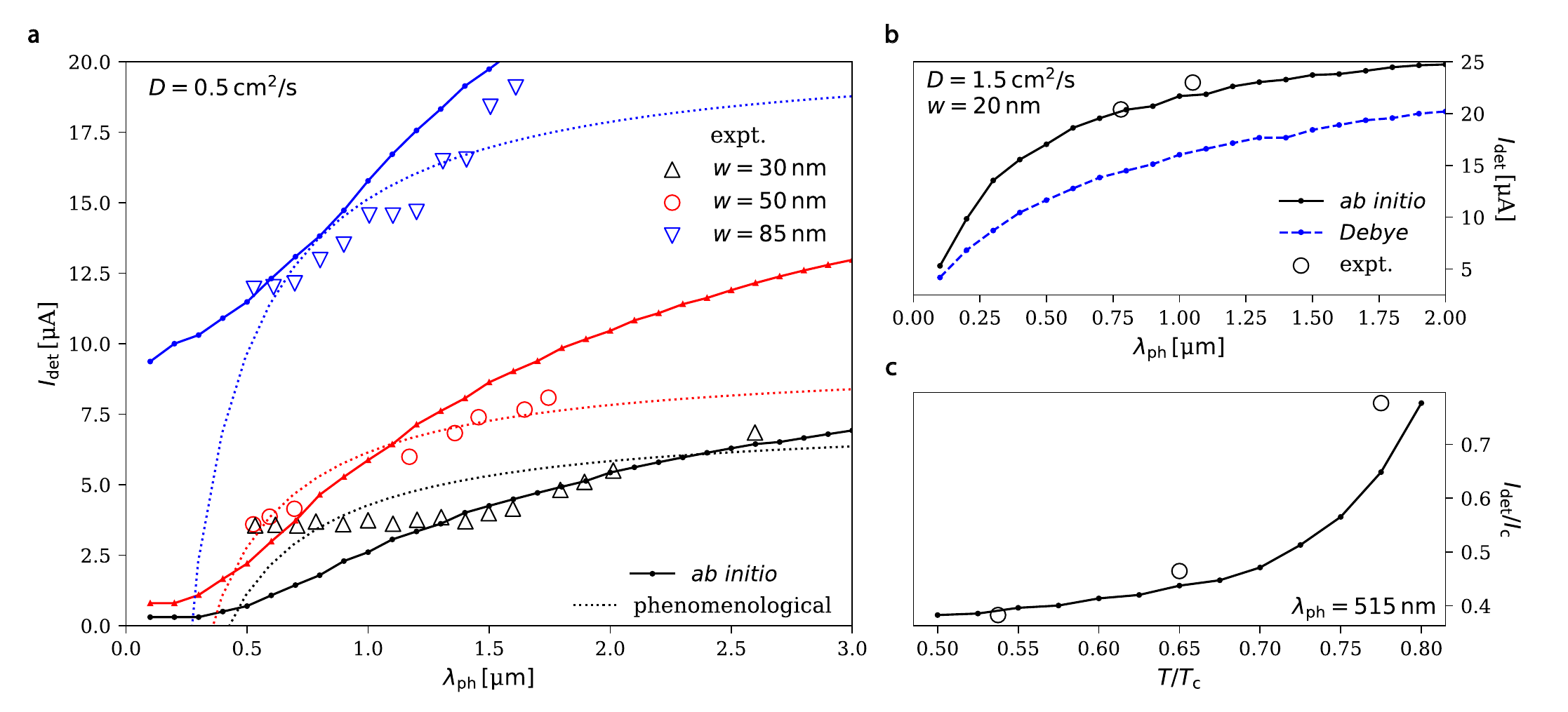}
    \caption{A comparison of the \textit{ab initio} model (solid lines) against the phenomenological diffusive hotspot model \cite{Semenov2001, Semenov_2021} (dotted lines), Debye model (dashed lines), and experimental data (large open markers) \cite{marsili2012efficient, Gourgues:19, 10.1063/5.0018818}.  (a) Determination of the detection current $I_{\mathrm{det}}$ for a given photon wavelength $\lambda_{\mathrm{ph}}$ and wire widths of $w = 30\,\rm{nm},\,50\,\rm{nm},$ and $85\,\rm{nm}$ with $D=0.5\,\mathrm{cm^2/s}$ compared to the experimental data of Ref. \cite{marsili2012efficient}. A value of $\eta = 0.2$ for the diffusive hotspot model gives the best fit (See Ref. \cite{Semenov_2021} for a definition of $\eta$). (b) $I_{\mathrm{det}}$ as a function of $\lambda_{\mathrm{ph}}$ for $w=20\,\mathrm{nm}$ and parameters consistent with the material used in the experimental data of Ref. \cite{10.1063/5.0018818}. Results using both the full-bandwidth electron-phonon coupling (\textit{ab initio}) and Debye model are displayed. (c) \textit{Ab initio} predictions of the normalized detection current $I_{\mathrm{det}}/I_{\rm{c}}$ as a function of the reduced temperature $T/T_\mathrm{c}$ for $\lambda_{\mathrm{ph}} = 515\,\rm{nm}$ compared to experimental data for the temperature-dependence of $\mathrm{NbTiN}$ from Ref. \cite{Gourgues:19}. Since the data in this figure is not for $\mathrm{NbN}$, only qualitative agreement of the temperature dependence is expected.  Calculations from the \textit{ab initio} model are in arbitrary units. }
    \label{fig:detection}
\end{figure*}

The microscopic treatment above is only suitable to account for the local suppression of superconductivity within $V_{\mathrm{init}}$ for $t \sim \tau_{\Delta}$. To determine if the suppression is sufficient to create a normal strip across the film, we must examine the two-dimensional dynamics of $\Delta$, including quantum fluctuations of the phase. These fluctuations are critical to consider, as experimental and theoretical evidence suggests that detection in SNSPDs is assisted by vortex motion or phase-slippage \cite{Engel_2015, PhysRevB.85.014505}.
In this picture, the suppression of superconductivity induced by the photon lowers the barrier for a $2\pi$-phase-slip of $\Delta$. Several processes can lead to phase-slip events in nanowires, including (1) the passage of a single vortex across the wire; (2) a quantum or thermally activated phase-slip; or (3) a vortex/anti-vortex pair that unbinds due to the Magnus force from $I_{\mathrm{B}}$. We refer to these processes collectively as phase-slip events. Once a phase-slip event occurs, Joule heating due to the nonzero $I_{\mathrm{B}}$ can lead to thermal runaway, destroying superconductivity across the strip and resulting in a voltage spike corresponding to a detection event. We describe our calculation of the phase-slip rate using the time-dependent Ginzburg-Landau (TDGL) equation in the Appendix. 
 
In Fig. \ref{fig:detection}a-c, we show the resulting dependence of $I_{\mathrm{det}}$ on $\lambda_{\mathrm{ph}}$. We note in comparing our model against $I_{\mathrm{det}}$, which corresponds to the onset of the plateau (90\%) in the photon count rate (PCR) curve, we have implicitly assumed that spatial inhomogeneities and thermal fluctuations dominate the shape of the PCR curve in $\mathrm{NbN}$. If Fano fluctuations were the dominant fluctuations, we would expect our model instead to provide predictions closer to the inflection point of the PCR curve \cite{PhysRevB.96.054507}, and the experimental data in Fig. \ref{fig:detection}a-c should be shifted downward correspondingly. We compare our calculations with the phenomenological diffusive normal-core model \cite{Semenov2001, Semenov_2021}, Debye model, and experimental data from Ref. \cite{marsili2012efficient, 10.1063/5.0018818}. We also check for qualitative agreement of the temperature dependence of $I_{\mathrm{det}}$ with the experimental data for $\mathrm{NbTiN}$ of Ref. \cite{Gourgues:19}. In our calculations, we assumed $D=0.5\,\mathrm{cm^2/s}$ for the polycrystalline detectors in Ref. \cite{marsili2012efficient} and $D=1.5\,\mathrm{cm^2/s}$ for the epitaxial detector of Ref. \cite{10.1063/5.0018818}. We observe that the qualitative behavior of the \textit{ab initio} model is similar to the experimental data, and there is reasonable quantitative agreement. The \textit{ab initio} model also improves over the predictions of the diffusive normal-core and Debye models. We emphasize that our model achieves this improved agreement without free parameters, which affirms the merit of our approach. We therefore propose that the methods presented here can be used to design the next generation of SNSPDs by enabling the exploration of new materials and geometries that extend device metrics to new regimes. 

As discussed in the Appendix, our TDGL formulation can be further improved and generalized to improve quantitative accuracy, and spatial inhomogeneity and thermal and Fano fluctuations can be incorporated to allow for the determination of the internal detection efficiency below $I_{\mathrm{det}}$, timing jitter, and the prediction of dark count rates \cite{PhysRevApplied.18.014006, 
PhysRevB.81.024502, PhysRevLett.107.137004} 
Optical absorption efficiency can also be calculated \cite{Sunter:18, Csete2013, Anant:08} and integrated to provide a complete end-to-end model from the crystal structure of a material to photon absorption and the resulting voltage spike and macroscopic circuit dynamics. 

\section{Quasiparticle-induced decoherence in transmon qubits}

The transmon qubit is currently a leading platform in the race to develop a practical quantum computer \cite{google}. However, since quantum computation is inherently error-prone, useful fault-tolerant quantum computation (FTQC) requires the implementation of error-correcting schemes. These error-correcting algorithms require high-fidelity physical qubits with long coherence times $T_1$. Through careful engineering of the design and material platform, transmon qubits have achieved coherence times on the order of hundreds of microseconds. However, this remains below the millisecond-scale coherence times typically desired for practical fault-tolerant quantum computing (FTQC). Further improvements necessitate an improved understanding of the fundamental sources of decoherence and their effects on qubit performance to motivate further engineering. 

\begin{figure*}[t]
    \centering
    \includegraphics[width=\linewidth]{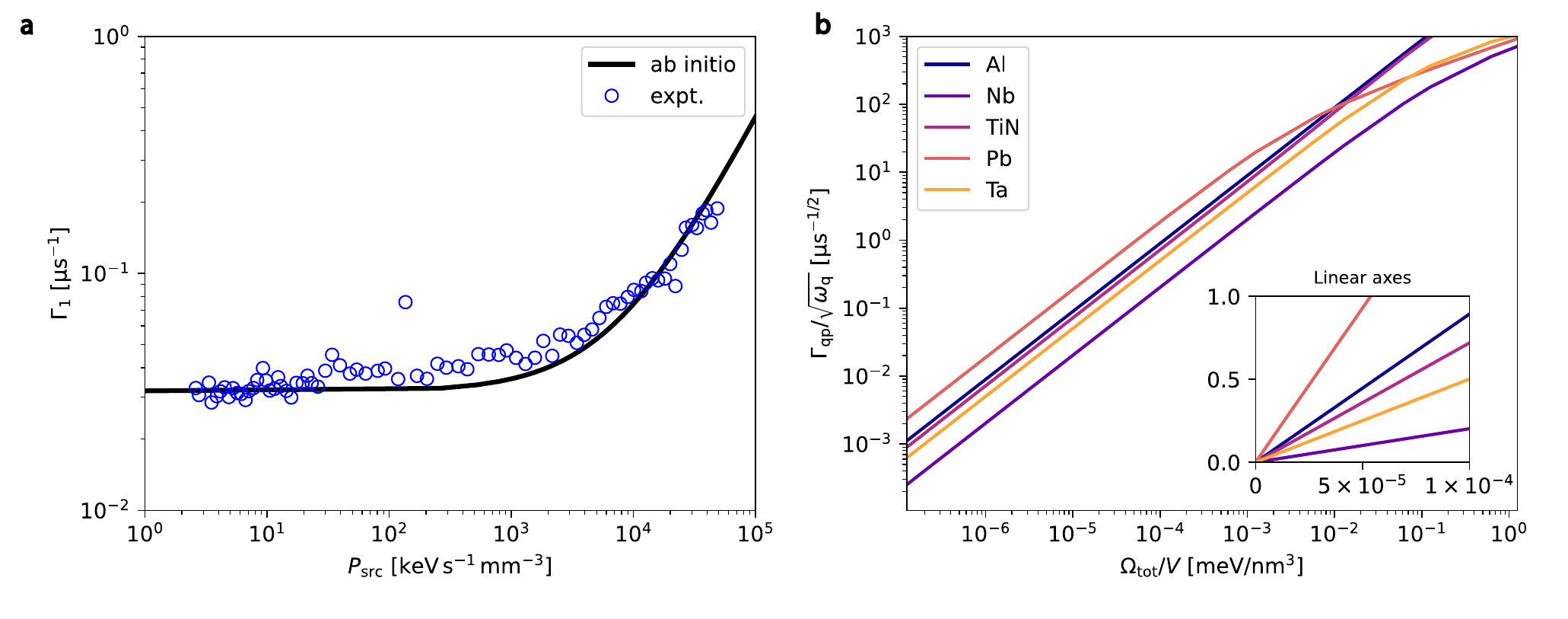}
    \caption{The energy-relaxation rate $\Gamma_1 = 1/T_1$ of a transmon qubit. (a) $\Gamma_1$ as a function of the radiation power density of a $\rm{Cu}^{64}$ source as described in Ref. \cite{Will-Oliver}. The \textit{ab initio} prediction is depicted with the experimental data. (b) Predictions from the \textit{ab initio} model of the dependence of $\Gamma_{\rm{qp}}$ on absorbed phonon energy density $\Omega_{\rm{tot}}/V$. Aluminum is found to be the most susceptible to quasiparticle poisoning, whereas niobium is the least.} 
    \label{fig:qubit}
\end{figure*}

Experimental data suggest that nonequilibrium quasiparticles play a role in correlated noise and decreasing $T_1$ in superconducting qubits \cite{Will-Oliver, Devoret}. Indeed, measurements of the quasiparticle densities in superconducting qubits have suggested that populations of nonequilibrium quasiparticles do not follow a thermal distribution near millikelvin temperatures ($\lesssim 150\,\rm{mK}$ in Al) due to an excess of quasiparticles \cite{Devoret}. These nonequilibrium populations have been attributed to the sensitivity of superconducting qubits to environmental radiation, including cosmic rays and the presence of naturally embedded radioactive isotopes in the environment \cite{Will-Oliver}. This radiation is either absorbed in the substrate, and the energy is transferred to the qubit via phonons originating in the substrate, or it can be directly absorbed by the Cooper pairs in the qubit. This process is depicted in Fig. \ref{fig:high-level}c-d.

The impact of nonequilibrium quasiparticles on $T_1$ is described by
\begin{equation}
\begin{aligned}
\label{eq:Gamma1}
    \Gamma_1 =\frac{1}{T_1} = \Gamma_{\rm{qp}} + \Gamma_{\rm{other}},
\end{aligned}
\end{equation} 
where $\Gamma_{\rm{qp}}$ is the energy-relaxation rate due to nonequilibrium quasiparticles and $\Gamma_{\rm{other}}$ encompasses all other loss mechanisms, including dielectric and two-level system losses. For a transmon, $\Gamma_{\rm{qp}}$ is given by 
\begin{equation}
\label{eq:QP-Rate}
    \Gamma_{\rm{qp}} = \sqrt{\frac{2\omega_{\rm{q}} |\Delta|}{\pi^2\hbar}}x_{\rm{qp}},
\end{equation} where $\omega_{\rm{q}}$ is the qubit frequency and $x_{\rm{qp}} = n/n_{\rm{cp}}$ is the density of quasiparticles $n$ normalized to the density of Cooper pairs in the material $n_{\rm{cp}}$ \cite{Will-Oliver, Geant4}. As expected, Eqs. \eqref{eq:Gamma1} and \eqref{eq:QP-Rate} suggest that $\Gamma_1$ can be reduced by decreasing the quasiparticle density $x_{\rm{qp}}$. To do so requires engineering the qubit or its substrate to reduce the rate of phonons transferred to the qubit or exploring alternative material platforms that may inherently possess a weaker response to environmental radiation. In both cases, a quantitatively accurate model of the qubit response to perturbation is desirable. 

Existing efforts to understand the qubit response to quasiparticle poisoning have relied on the quasiparticle dynamics provided by the Rothwarf-Taylor model \cite{Will-Oliver, Geant4}, which is a phenomenological description of quasiparticle recombination, scattering, and generation\footnote{The Rothwarf-Taylor equations can be derived from Eqs. \eqref{eq:qp-dist} and \eqref{eq:ph-dist} \cite{chang_kinetic-equation_1977}.}. These models lack a rigorous description of these rates and rely on fitting to experimental data, hence limiting their applicability to exploring novel materials or physics. Our model is well-suited to fill this gap since it can predict these rates from more fundamental material properties, enabling the comparison of quasiparticle dynamics across a wide range of materials. 

In Fig. \ref{fig:qubit}a, we present calculations of $\Gamma_1$ for an Al-based transmon using the DFPT calculations shown in Fig. \ref{fig:DFT}c. $\Gamma_1$ is calculated for different radiation source power densities $P_{\rm{src}}$, allowing us to compare our model with the experimental data of Ref. \cite{Will-Oliver}. To determine $\Gamma_1$ from our model, we injected an initial phonon distribution with a modified phonon bubble initial condition as detailed in the Appendix. An energy $\chi P_{\rm{src}}$ was used for the phonon bubble, where $\chi = 3.5\times10^{4} \,\rm{s}$ determines the percentage of energy deposited into the qubit from phonons in the substrate. A rigorous determination of $\chi$ is possible through a detailed accounting of the phonon dynamics in the substrate in a simulation package such as Geant4 \cite{Geant4}. For each energy, we solved Eqs. \eqref{eq:qp-dist} and \eqref{eq:ph-dist} for the thermalized quasiparticle distribution and extracted the resulting quasiparticle density. Eq. \eqref{eq:QP-Rate} was then used to compute $\Gamma_{\rm{qp}}$ for each $P_{\rm{src}}$. $\Gamma_{\rm{other}}$ was determined by averaging $\Gamma_1$ over the 10 lowest $P_{\rm{src}}$ experimental data points. As evidenced by Fig. \ref{fig:qubit}a, we obtained good agreement using our approach. 


To compare the susceptibility of different materials to ambient radiation, we calculated $\Gamma_{\rm{qp}}$ for Al, Nb, TiN, Ta, and Pb for different injected phonon energy densities $\Omega_{\rm{tot}}/V$. The material parameters used in the simulations are listed in the Appendix, and parameters not unique to the materials were kept consistent in the simulations. In Fig. \ref{fig:qubit}b, the results of the calculations for each material are displayed for direct comparison. Al was found to be the most susceptible to quasiparticle poisoning, which is unsurprising because of its relatively small superconducting gap. At $\Omega_{\rm{tot}}/V$, $\Gamma_{\rm{qp}}$ exhibits a similar dependence for all materials; however, the rates of quasiparticle generation in different materials diverge at larger $\Omega_{\rm{tot}}/V$, with Nb, Pb, and Ta increasing at a slower rate than Al and TiN. Interestingly, our \textit{ab initio} model predicts a low susceptibility to quasiparticle poisoning in Ta that is on par with larger gap superconductors like Nb and Pb. This result may in part explain the reduced quasiparticle tunneling rate and longer coherence times observed in Ta transmon qubits \cite{tennant2022low, Princeton}.  

\section{Conclusions \& Outlook}

In summary, we have demonstrated a framework for modeling the performance of superconducting devices \textit{ab initio}. To illustrate the effectiveness of our approach, we applied our model to describe the detection mechanism of SNSPDs and quasiparticle poisoning of superconducting qubits. We demonstrate that our model improves significantly over existing phenomenological models of quasiparticle dynamics while using no free parameters. We emphasize that the methods discussed here can be extended to other superconducting devices and materials. Future work may investigate an even larger assortment of materials and devices to search for materials that perform beyond the current state-of-the-art. Recently, there has been an effort to develop \textit{ab initio} methods for searching for high-$T_{\rm{c}}$ superconductors \cite{Pellegrini2024}. The methods described here may enable a similar systematic search for materials that optimize the performance of superconducting devices. 


\section{acknowledgments}
    The authors are deeply grateful to Rohit Prasankumar and Mukund Vengalattore for their helpful discussions. The authors are also grateful to Jason Allmaras and Phillip Donald Keathley for their insightful feedback on this manuscript. This work was funded in part by the Defense Sciences Office (DSO) of the Defense Advanced Research Projects Agency (DARPA) (HR0011-24-9-0311). AS and EB acknowledge support from the NSF GRFP, RF acknowledges support from the Alan McWhorter fellowship, OM acknowledges support from the NDSEG fellowship. MS and CH acknowledge support from Intellectual Ventures Property Holdings and usage of computational resources of the dCluster of the Graz University of Technology.

\section{Appendix}

\subsection{Numerical solutions to the kinetic equations}

The kernel functions in Eqs. \eqref{eq:qp-dist} and \eqref{eq:ph-dist} of the main text are given by
\begin{subequations}
\begin{widetext}
\begin{equation}
\begin{aligned}
    K_{ph-\mathrm{e}} (E,\Omega) & = \left( 1 - \frac{\Delta^2}{E(E+\Omega)} \right) \Bigg\{ f(E) [1-f(E+\Omega)]n(\Omega) - f(E+\Omega)[1-f(E)][n(\Omega)+1]\Bigg\}
\end{aligned}
\end{equation}
\begin{equation}
\begin{aligned}
    K_{e-\mathrm{ph}}(E,\Omega) & = \left( 1 - \frac{\Delta^2}{E(E-\Omega)} \right) \Bigg\{ f(E) [1-f(E-\Omega)][n(\Omega)+1] - [1-f(E)]f(E-\Omega)n(\Omega) \Bigg\}
\end{aligned}
\end{equation}
\begin{equation}
\begin{aligned}
     K_{\mathrm{R}}(E,\Omega) = \left( 1 + \frac{\Delta^2}{E(\Omega-E)} \right) \Bigg\{ f(E)f(\Omega-E)[n(\Omega)+1] - [1-f(E)][1-f(\Omega-E)]n(\Omega)\Bigg\}
\end{aligned}
\end{equation}
\begin{equation}
\begin{aligned}
     K_{\mathrm{S}}(E, E', \Omega) = \left(1-\frac{\Delta^2}{E E'} \right) \Bigg\{ f(E)[1-f(E')]n(\Omega) - f(E')[1-f(E)][n(\Omega) + 1] \Bigg\}
\end{aligned}
\end{equation}
\begin{equation}
\begin{aligned}
     K_{\mathrm{B}}(E, E', \Omega) = \frac{1}{2}\left( 1 + \frac{\Delta^2}{E E'} \right) \Bigg\{ [1-f(E)][1-f(E')]n(\Omega) - f(E)f(E')[n(\Omega) + 1] \Bigg\}.
\end{aligned}
\end{equation}
\end{widetext}
\end{subequations} Numerical solutions to Eq. \eqref{eq:qp-dist}, \eqref{eq:ph-dist}, \and (5a-e) were obtained with a forward Euler scheme, where the integrals were evaluated numerically with a trapezoidal method at each timestep. Where applicable, vectorized operations were used to speed up the computation. We introduced a cutoff above the maximum acoustic phonon energy for the integral bounds in Eqs. \eqref{eq:qp-dist} and \eqref{eq:ph-dist} while ensuring that the numerical solutions were not affected by the choice of a larger cutoff and remained below our error criteria. As discussed in Appendix \ref{appendix:Usadel}, Eqs. \eqref{eqn:usadel_diff} and \eqref{eqn:usadel_sc} are solved self-consistently for a given $I_{\mathrm{B}}$ to determine $\rho(E)$ and $|\Delta| = |\Delta(I_\mathrm{B}, T)|$. Details on obtaining the numerical solutions to the Usadel equation are discussed in Appendix \ref{appendix:Usadel}. At each time step, it was checked that the total energy of the electronic system 
\begin{equation}
\begin{aligned}
    E_e = & 4 N(0)  V_{\mathrm{init}} \Bigg\{ \int_{|\Delta|}^{\infty}  d\!E \, E \rho(E) f(E) \\ & - \frac{|\Delta|^2}{4} \left(\frac{1}{2} + \ln{\frac{\Delta_0}{|\Delta|}} \right) \Bigg\}
\end{aligned}
\end{equation}
and the phonon system
\begin{equation}
    E_{\mathrm{vib}} = V_{\mathrm{init}} \int_0^{\infty}  d\Omega \, \Omega N F(\Omega)n(\Omega)
\end{equation}
is conserved within 0.1\% of the starting energy for $t \leq \tau_{\Delta}$ when $\tau_{\mathrm{esc}} = \infty$. In most of our simulations, this error criterion was often exceeded by orders of magnitude. It was also checked that increasing the number of discrete energies and time steps used in the numerical solutions did not change the calculation results significantly. We found using a grid of energies spaced by $\sim \Delta/15$ and a grid of times spaced by $10\,\rm{fs}$ sufficient to meet our error criteria by approximately an order of magnitude in most cases. For the simulations of optical irradiation in NbN, with $\tau_{\mathrm{esc}} = 10\,\mathrm{ps}$, about $\sim 25\%$ of the photon energy was lost to the substrate.  

\begin{figure*}[t]
    \centering
    \includegraphics[width=\textwidth]{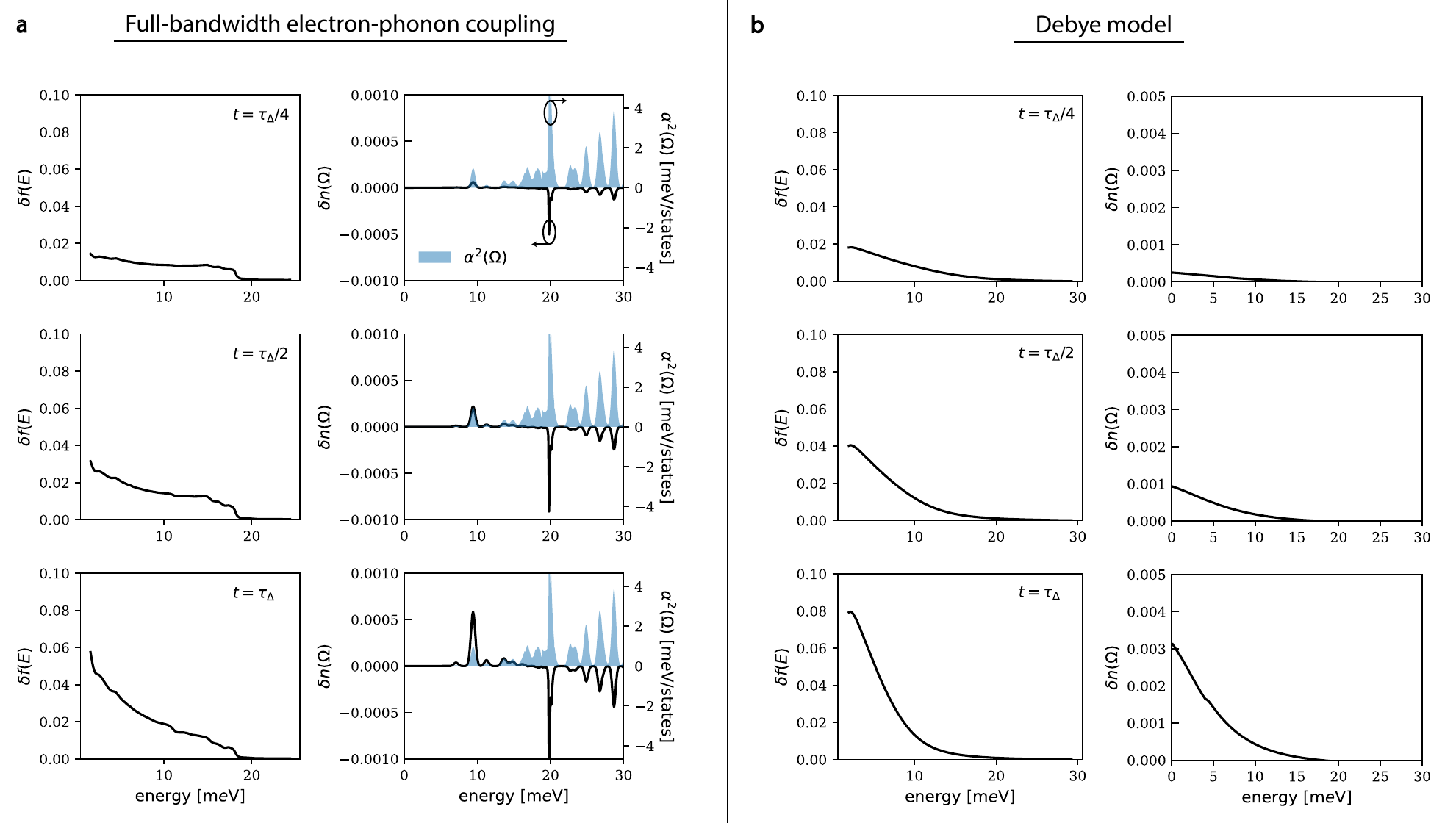}
    \caption{Evolution of the excess quasiparticle $\delta f(E)$ and phonon $\delta n(\Omega)$ distributions for $t = \tau_{\Delta}/4,\,\tau_{\Delta}/2,$ and $\tau_\Delta$. (a) Calculations using the full-bandwidth electron-phonon coupling and phonon density of states as determined by DFPT calculations. In these figures, $\alpha^2(\Omega)$ is plotted alongside $\delta n(\Omega)$ to demonstrate the role of the electron-phonon coupling in determining the evolution of the phonon distribution. (b) The evolution of $\delta f(E)$ and $\delta n(\Omega)$ assuming a Debye approximation for the electron-phonon coupling and phonon density of states. }
    \label{fig:time-dependent-evolution}
\end{figure*}

To determine the phonon-bubble initial condition for the SNSPD simulations, we use Eq. \eqref{eq:ph-dist} and approximate the initial excess quasiparticle distribution $\delta\!f(E) = f(E) - f^{\mathrm{eq}}(E)$ as a delta function centered at $E_\lambda$, which gives that the initial phonon population is $n_0(\Omega) = \beta(E_\lambda) \alpha^2(\Omega)$ \cite{chang_kinetic-equation_1977}, where the parameter $\beta(E_\lambda)$ ensures that the initial energy of the phonon system is equal to the photon energy. Thus, $n_0(\Omega)$ and an equilibrium quasiparticle (Fermi-Dirac) distribution $f^{\mathrm{eq}}(E)$ characterize the phonon bubble. Eqs. \eqref{eq:qp-dist} and \eqref{eq:ph-dist} with the phonon-bubble initial condition then provide the subsequent quasiparticle and phonon dynamics that result from the absorption of a photon. In principle, one could instead use an initial condition with a delta function centered at $E_\lambda$ for $f(E)$ and an equilibrium distribution for $n(\Omega)$. However, a solution over the full range of energies up to $E_\lambda$ is a computationally expensive task, and the initial phonon bubble allows us to significantly reduce the computational burden. 

In Fig. \ref{fig:time-dependent-evolution}a, we plot the time evolution of the excess quasiparticle $\delta f(E)$ and phonon distributions $\delta n(\Omega)$ resulting from the ``phonon bubble'' initial condition in NbN when considering the full-bandwidth electron-phonon coupling and phonon density of states as computed with DFPT calculations. The $\delta f(E)$ and $\delta n(\Omega)$ resulting from the Debye approximation are also plotted in Fig. \ref{fig:time-dependent-evolution}b for comparison with the results obtained with the exact from of $\alpha^2F(\Omega)$ and $F(\Omega)$. The qualitative behavior between the solutions of the kinetic equations with the DFPT and Debye models is similar. However, the electron-phonon coupling leads to non-trivial changes in the generated quasiparticle and phonon populations. We also note that the discontinuity in the phonon distribution near $\Omega = 2|\Delta|$ is more prominent with the Debye model. However, unlike in Ref. \cite{chang_kinetic-equation_1977}, the $2|\Delta|$ discontinuity is smeared since we incorporate the effects of the finite $\rho(E)$ resulting from a nonzero bias current. For NbN, it was found that the Debye model generally overestimates the rate of quasiparticle generation, which is reflected in Fig. \ref{fig:detection}b of the main text, where it can be seen that the prediction of $I_{\mathrm{det}}$ from the Debye model is consistently below the experimental data and predictions of the \textit{ab initio} model.

For the transmon simulations, an initial volume of $V_{\rm{init}} = d^3$ was chosen as it is a natural length scale of the device that is consistent across every material. Since we are most interested in the energy density deposited into the superconductor, which can be computed from Geant4 simulations, this choice is appropriate to describe dynamics.

\subsection{Time-dependent Ginzburg-Landau equation}

\begin{figure*}[t]
    \centering
    \includegraphics[width=\linewidth]{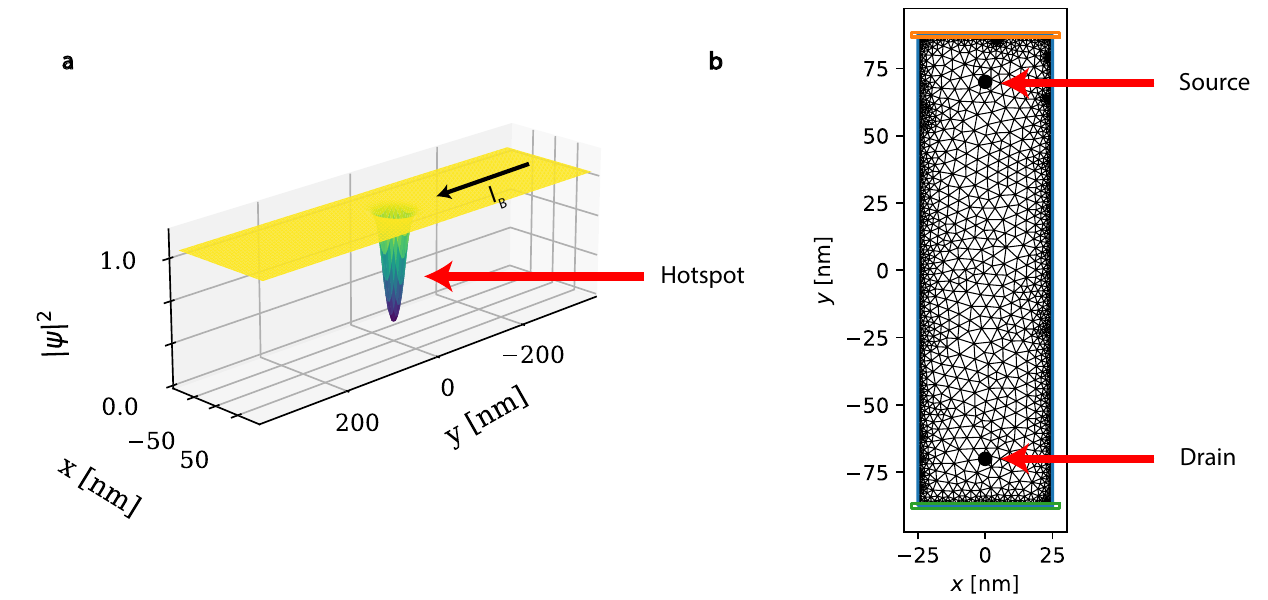}
    \caption{Time-dependent Ginzburg-Landau simulation of the suppression of the normalized superconducting order parameter $\psi$ following the absorption of a photon. (a) Magnitude of $|\psi|^2$ after photon absorption. The photon-induced hotspot is labeled with a red arrow, and the direction of the bias current $I_{\rm B}$ from source to drain is indicated by the black arrow. (b) Meshing used in pyTDGL for numerical solutions of the TDGL equations for a wire width of $w = 50\,\rm{nm}$. The source and drain for the current are labeled with red arrows.}
    \label{fig:TDGL-mesh}
\end{figure*}

As discussed in the main text, we connect the results for the microscopic dynamics of the quasiparticle and phonon systems to the mesoscopic dynamics of $\Delta$ via the time-dependent Ginzburg-Landau (TDGL) equation. This is useful in the case of modeling optical irradiation since phase-slip dynamics are naturally captured by the TDGL equation \begin{equation}
\label{eq:tdgl}
\begin{aligned}
    \frac{u}{\sqrt{1+\gamma^2 |\psi|^2}} & \left( \frac{\partial}{\partial t} + i \mu + \frac{1}{2} \gamma^2 \frac{\partial |\psi|^2}{\partial t} \right) \psi = \\ & \xi^2 ( \nabla - 2 i e \mathbf{A})^2\psi + (\alpha-|\psi|^2)\psi,
\end{aligned}
\end{equation} where we have used dimensionless units with $u = \pi^4/14\zeta(3) = 5.79$, $\gamma = 10$, the normalized superconducting order parameter $\psi = \Delta/|\Delta_0|$, and the electric scalar potential $\mu(\mathbf{r}, t)$ which satisfies a Poisson equation \cite{PhysRevLett.40.1041}. The parameter \begin{equation}
    \alpha(\mathbf{r}, t) = (1 - T/T_{\mathrm{c}} - \varepsilon(\mathbf{r}, t))/(1 - T/T_{\mathrm{c}})
\end{equation} models the photon-induced suppression of $\psi$ at position $\mathbf{r}$ and time $t$ ($\alpha = 1$ at equilibrium), where $\varepsilon(\mathbf{r}, t)$ is calculated from the microscopic dynamics using Eq. \eqref{eq:epsilon}. Here, we assume that the hotspot grows isotropically with a time-dependent radius $|\mathbf{r}(t)| = \sqrt{D t}$. A more rigorous treatment of quasiparticle and phonon diffusion, along with solving Eqs. \eqref{eq:qp-dist} and \eqref{eq:ph-dist} self-consistently with Eq. \eqref{eq:real-axis-sc}, would significantly improve the accuracy for wires with larger $w$, which possess longer latency times between absorption and detection. We also note that $\tau_\mathrm{th} > \tau_{\Delta}$ implies the local electron and phonon temperatures are still evolving when the region of suppressed superconductivity has diffused beyond $V_{\mathrm{init}}$. Hence, caution must be exercised when assuming a well-defined electron and phonon temperature exists during the early stages of the quasiparticle cascade; however, a rigorous treatment of the microscopic diffusion could help address this issue. The generalized TDGL equation with Usadel corrections for the supercurrent density $j_{\mathrm{S}}$ and $\Delta$ can also be used in place of Eq. \eqref{eq:tdgl} to improve validity at lower temperatures and large deviations from equilibrium \cite{vodolazov_single-photon_2017}. 
Incorporating thermal equations for the electron and phonon temperatures and a circuit model to account for Joule heating, the kinetic inductance of the film, and the external circuitry will further improve the quantitative accuracy. These additional equations are most relevant in the presence of a small shunt resistance or small $I_{\mathrm{B}}$, where these processes may play a role in the initial phase slip and in initiating thermal runaway. This would likely improve the quantitative agreement at shorter $\lambda_{\mathrm{ph}}$ for $w = 30\,\rm{nm}$ and $D=0.5\,\mathrm{cm^2/s}$ in Fig. \ref{fig:suppression}d. However, as presented, the current model is sufficient to obtain reasonable quantitative accuracy.

We solve Eq. \eqref{eq:tdgl} with the Python package pyTDGL \cite{Bishop-Van_Horn2023-wr}. In these simulations, for each photon wavelength $\lambda_{\mathrm{ph}}$, we varied $I_\mathrm{B}$ while checking if the voltage arising from the formation of a normal strip across the wire exceeded a threshold and the phase difference across the terminals of the device exceeded $2\pi$. The current at which these criteria were met was determined to be the detection current $I_{\mathrm{det}}$. Note that since we perform the simulation over a coarse grid of bias currents, some quantization error is introduced. We restricted our simulation time to $t < 15\,\mathrm{ps}$ for the wires with $w=30\,\rm{nm}$ and $w=50\,\rm{nm}$ and $t < 25\,\mathrm{ps}$ for $w=85\,\rm{nm}$ to account for the longer latency between photon absorption and detection. For the epitaxial detector, where diffusion was much faster, with $w=20\,\rm{nm}$ the simulation time was restricted to $t < 5\,\rm{ps}$. Material parameters were kept consistent between the microscopic and TDGL simulations. In the simulations for $w=20\,\rm{nm}$, $30\,\rm{nm}$, and $50\,\rm{nm}$, the max edge length of the two-dimensional mesh was set to $2 \xi_{\rm{c}}$. For $w=85\,\rm{nm}$, the max edge length was set to $2.5\xi_{\rm{c}}$ to decrease runtime. Examples of the meshing are displayed in Fig. \ref{fig:TDGL-mesh}. By default, the pyTDGL package only allows for simulations down to $T = T_{\rm{c}} / 2$, so to simulate the behavior of the experimental data for SNSPDs in Ref. \cite{10.1063/5.0018818} and \cite{marsili2012efficient}, where the SNSPDs were measured at $T \approx 0.2 T_{\rm{c}}$, we adjusted the zero temperature values of $\xi_{\rm{c}}$ and $\lambda_{\rm{L}}$ using their temperature dependencies from TDGL to make them consistent with the values at $T = 0.2 T_{\rm{c}}$.

\subsection{The Usadel equation}
\label{appendix:Usadel}

\begin{figure*}[t]
    \centering
    \includegraphics[width=\linewidth]{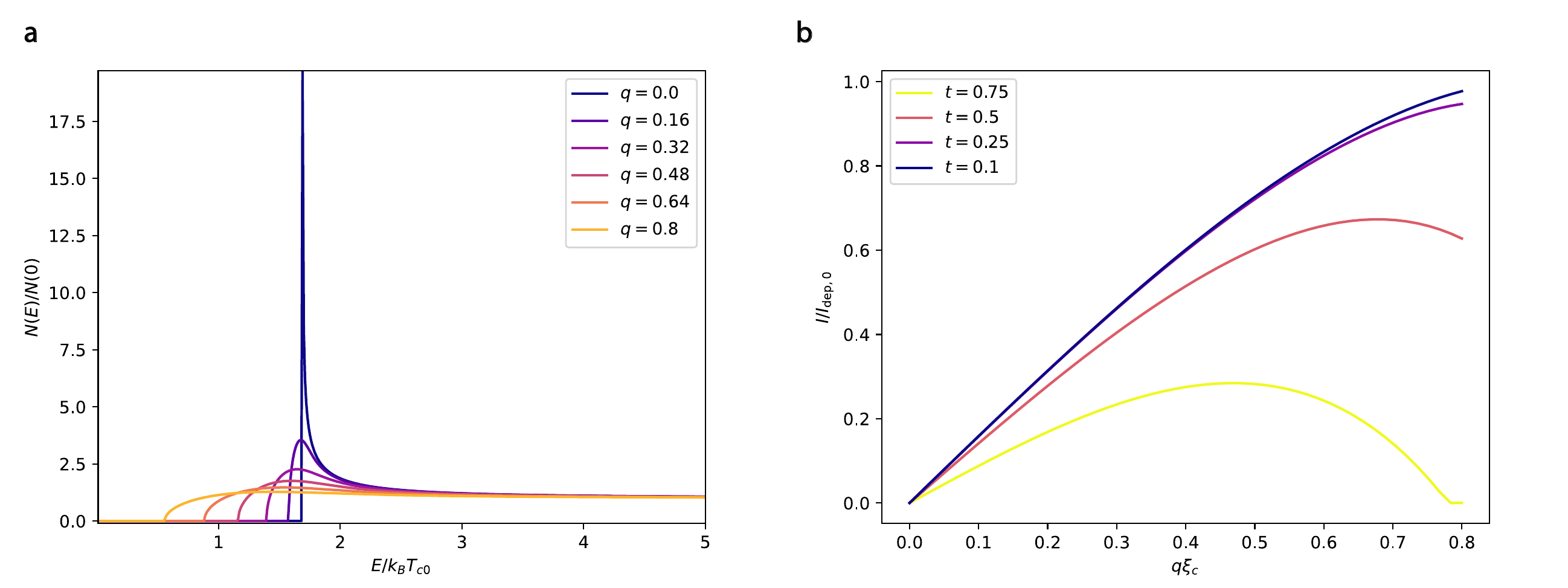}
    \caption{(a) The normalized quasiparticle density of states $\rho(E) = N(E)/N(0)$ for several values of the superfluid momentum $q$ at $T \sim 0\,\rm{K}$. (b) The supercurrent $I = j/wd$ normalized to the depairing current at zero temperature $I/I_{\mathrm{dep,0}}$ for several values of the reduced temperature $t = T/T_\mathrm{c}$.}
    \label{fig:QPDOS}
\end{figure*}

The normalized quasiparticle density of states for a superconducting film in the dirty limit is given by $\rho(E) = \mathrm{Re}\{ \cos \Theta (E) \}$, where $\Theta(E)$ can be obtained from the Usadel equation on the imaginary-frequency axis
\begin{equation}
\label{eqn:usadel_diff}
\begin{aligned}
    & \hbar D \nabla^2\Theta(\vec{\mathbf{r}},i \omega_n) + 2|\Delta(\vec{\mathbf{r}})|\cos\Theta(\vec{\mathbf{r}},i \omega_n) \\ & - \left(2\hbar\omega_n + \frac{D}{\hbar}q^2\cos\Theta(\vec{\mathbf{r}},i \omega_n)\right)\sin\Theta(\vec{\mathbf{r}},i \omega_n)  = 0
\end{aligned}
\end{equation} 
and performing an analytical continuation to the real-frequency axis $i \omega_n \rightarrow E + i0^+$.  For a uniform film, the spatial dependence of $\Theta$ and $|\Delta|$ can be ignored, there are no central boundary conditions, and the diffusive term $\hbar D\nabla^2\Theta(\vec{\mathbf{r}},i\omega_n)$ is zero.
The order parameter $|\Delta|$ satisfies the BCS self-consistency equation
\begin{equation}
    \label{eqn:usadel_sc}
  |\Delta| \ln\left(\frac{T}{T_{{\rm c}}}\right) + 2\pi k_{\mathrm{B}} T \sum_{\omega_n \geq 0}\left(\frac{|\Delta|}{\hbar\omega_n} - \sin\Theta(i \omega_n)\right) = 0,
\end{equation}
where $k_\mathrm{B}$ is the Boltzmann constant, $T_{\mathrm{c}}$ is the critical temperature, $\omega_n =  \pi T (2n+1)k_B/\hbar$ is the $n$-th Matsubara frequency, and $\Theta(i \omega_n)$ is the pairing-angle parametrization of the Nambu-Gor'kov Green's function \cite{usadel1970generalized, PhysRevB.36.5665}. The full Migdal-Eliashberg self-consistency equations on the real-frequency axis, along with the strong-coupling Usadel equation \cite{PhysRevB.36.5665}, could be used instead of the BCS self-consistency equation (Eqs. \eqref{eq:real-axis-sc} and \eqref{eqn:usadel_sc}); however, the resulting complexity and effort to solve these equations would have been significant and beyond the scope of the current work. This approximation limits the quantitative accuracy of our model. The superfluid momentum $q$ is related to the supercurrent density via 
\begin{equation} 
\label{eq:current}
\begin{aligned}
    j_\mathrm{S} = \frac{I_{\mathrm{B}}}{wd} & = \frac{2 \pi k_{\mathrm{B}} T}{|e|}\sigma_0 q \sum_{n=0}^{\infty} \sin^2\Theta(i \omega_n),
\end{aligned}
\end{equation} 
where $\sigma_0 = 1/\rho_{\mathrm{N}}$ is the Drude conductivity. We define the spectral function \begin{equation}
R(E,\Delta) = \mathrm{Im}\{ \sin \Theta(E) \}\end{equation} from the main text.

Eqs. \eqref{eqn:usadel_diff} and \eqref{eqn:usadel_sc} are solved simultaneously with a left-preconditioned Newton-Krylov method \cite{knoll_jacobian-free_2004, virtanen2025magnetoelectric}. The preconditioner is constructed by sparsifying the Jacobian of Eqs. \eqref{eqn:usadel_diff} and \eqref{eqn:usadel_sc} assuming a BCS ($q = 0$) solution. The infinite Matsubara sum in Eq. \eqref{eqn:usadel_sc} is approximated with a quadrature rule for sums \cite{monien_gaussian_2016}.

A nonzero bias current induces a smearing of the singularity at the gap-edge $\Delta$ in the quasiparticle density of states $\rho(E)$. As discussed above, for simplicity, we solve the Usadel equation in the weak-coupling limit, which differs from the strong-coupling form by a scaling factor in the self-consistency equation, an additional self-consistency equation for the wave-function renormalization parameter, and a frequency-dependent order parameter $\Delta(i \omega_n)$. In the strong-coupling case, these equations are more difficult to solve, but the relative shape of $\rho(E)$ is qualitatively similar. To better match the experimental data for $\mathrm{NbN}$, we scale the results of the weak-coupling theory by a factor of $2.1/1.76$ to account for the strong-coupling nature of NbN and retain the expected relationship between $\Delta$ and $T_{\rm{c}}$ ($\Delta_0 = 2.1 k_\mathrm{B} T_\mathrm{c}$) as predicted by Migdal-Eliashberg theory and observed experimentally \cite{Semenov2001}. For bias currents on the order of the depairing current, the smearing takes on a similar form as seen in Fig. \ref{fig:QPDOS}a for different values of the superfluid momentum $q$. The superfluid momentum $q$ is connected to the supercurrent via Eq. \eqref{eq:current}, and its solutions are shown in Fig. \ref{fig:QPDOS}b for several different reduced temperatures $t = T/T_{\rm c}$. The value of $I$ at which $dI/dq = 0$ is the temperature-dependent depairing current $I_{\mathrm{dep}}$. In our solutions, we found that for $0.5 I_{\mathrm{dep}} \leq I_{\mathrm{B}} \leq 0.9 I_{\mathrm{dep}}$, where $I_{\mathrm{dep}}$ is the depairing current, there was not a strong dependence of the generated quasiparticle population on $I_{\mathrm{B}}$. Hence, we set $I_{\mathrm{B}} = 0.5 I_{\mathrm{dep}}$, which incorporates the effect of smearing in $\rho(E)$ while also preserving the generality of the results to polycrystalline devices with switching currents on the order of $0.5 I_{\mathrm{dep}}$.

\subsection{Density functional theory calculations}
We employed the Quantum Espresso code \cite{giannozzi2017advanced} to compute the structural, electronic, and harmonic phonon properties of the bulk materials using first-principles DFPT. The exchange-correlation functional was described by the Perdew-Burke-Ernzerhof (PBE) \cite{perdew1996generalized} version of the generalized gradient approximation (GGA) in combination with optimized norm-conserving Vanderbilt pseudopotentials \cite{schlipf2015optimization, hamann2013optimized}. The convergence threshold for the self-consistent field was set to $10^{-10}$ Ry for the energy difference between consecutive electronic steps and structural relaxations were performed until the forces on each atom were less than $10^{-6}$ Ry/\AA. A plane-wave kinetic energy cutoff of 80 Ry was used for all materials except Ta, where 140 Ry was used, along with an $18\times18\times18$ k-point grid and a Methfessel-Paxton (MP) smearing \cite{methfessel1989high} of 0.02 Ry to sample the Brillouin zone for self-consistent calculations. For NbN, the MP smearing had to be set to 0.2 Ry to avoid imaginary frequencies in the harmonic DFPT calculation of phonon properties, as disorder, impurities, and anharmonic effects are not accounted for. As can be appreciated in Fig.~\ref{fig:DFT}, the calculated phonon dispersion compares well to experimental data~\cite{christensen1979phonon}. Spin-orbit coupling effects were included for Pb and Ta.

For phonon calculations, we employed a $6\times6\times6$ q-point grid ($8\times8\times8$ for Ta) and a threshold of self-consistency of $10^{-14}$ Ry to obtain the dynamical matrices within the harmonic approximation. To improve convergence for NbN and Pb, we interpolated the electronic and phononic properties onto finer grids utilizing the Electron-Phonon Wannier (EPW) code \cite{lee2023electron}. Specifically, interpolation to fine grids of $30\times30\times 30$ for both electrons and phonons was performed to obtain isotropic Eliashberg spectral functions $\alpha^2F(\Omega)$.

\subsection{Material Parameters}

\begin{table}[h]
    \centering
    \renewcommand{\arraystretch}{1.2}
    \begin{tabular}{l c c c c c c}
        \toprule
        \textbf{Symbol} & \textbf{NbN} & \textbf{Al} & \textbf{Nb} & \textbf{TiN} & \textbf{Pb} & \textbf{Ta} \\ \midrule
        $N~[\rm{nm}^3]$ & 48 & 60 & 55 & 37 & 33 & 56 \\
        $\rho_N~[\Omega\cdot\rm{nm}]$ & 1040 & 28 & 152 & 220 & 220 & 136 \\
        $E_{\rm{F}}~{[\rm{eV}]}$ & 5.895 & 11.7 & 5.32 & 4.2 & 10.87 & 5.32 \\
        $d~[\rm{nm}]$ & 5 & 200 & 200 & 200 & 200 & 200 \\
        $\tau_{\rm{esc}}~[\rm{ps}]$ & 15 & 100 & 100 & 100 & 100 & 100 \\
        $\Delta~[\rm{meV}]$ & 1.8 & 0.2 & 1.2 & 1.4 & 1.1 & 0.7 \\ 
        $T_c~[\rm{K}]$ & 10 & 1.3 & 8 & 9 & 7 & 4.2 \\ \bottomrule
    \end{tabular}
    \caption{Material parameters used in the simulations in the main text for NbN, Al, Nb, TiN, Pb, and Ta. $N$ is the ion density, $\rho_N$ is the bulk resistivity, $E_{\rm{F}}$ is the Fermi energy, $d$ is the device thickness for a superconducting nanowire single-photon detector for NbN and the thickness for a transmon qubit for Al, Nb, TiN, Pb, and Ta, $\tau_{\rm{esc}}$ is the corresponding characteristic time for phonon escape to the substrate, $\Delta = |\Delta|$ is the magnitude of the superconducting order parameter, and $T_{\rm{c}}$ is the critical temperature. For the Cooper pair density $n_{\rm{cp}}$, the quoted value in the literature of $0.004\,\rm{nm}^3$ was used for Al \cite{Will-Oliver}, and the rest were estimated using the London penetration depth. }
    \label{tab:material_parameters}
\end{table}

Table \ref{tab:material_parameters} shows the material parameters used in the simulations of the main text. These values are approximate and can also be obtained \textit{ab initio} from DFT rather than experimental data \cite{giustino2014materials, Pellegrini2024}. For NbN, $|\Delta|$ and $T_{\rm{c}}$ were set to resemble the expected values for a disordered polycrystalline film. The superconducting properties of all other materials were calculated from the BCS self-consistency equation and the DFPT calculations. 


\bibliography{apssamp}


\end{document}